\documentclass[journal]{IEEEtran}

\IEEEoverridecommandlockouts

\usepackage{amsmath,amssymb,amsfonts,empheq,bm}
\usepackage{algorithmic}
\usepackage{graphicx}
\usepackage{textcomp}
\usepackage{xcolor}
\usepackage{float}
\usepackage{setspace}
\usepackage{comment}

\usepackage{dotlessi}
\usepackage{layouts}
\usepackage{mathtools}
\usepackage{subcaption}
\usepackage[font=small]{caption}
\usepackage{multirow}
\usepackage{amsmath}
\usepackage{amsfonts}
\newcommand{\nx}{n_x}
\newcommand{\ny}{n_y}
\newcommand{\nz}{n_z}
\usepackage{bm}
\usepackage{etoolbox}
\usepackage{cite}

\makeatletter
\patchcmd{\@sect}{\MakeLowercase}{}{}{}
\makeatother

\newcommand{\bfg}[1]{\boldsymbol{#1}}

\begin{document}

\title{Synthetic Discrete Inertia} 

\author{%
  {\'A}ngel Vaca, \IEEEmembership{IEEE Student Member} %
  and Federico Milano, \IEEEmembership{IEEE Fellow}%
  \thanks{\'A.~Vaca and F.~Milano are with the School of Electrical and Electronic Engineering, University College Dublin, Belfield Campus, D04V1W8, Ireland. e-mails: angel.vaca1@ucdconnect.ie, federico.milano@ucd.ie}%
  \thanks{This work is supported by the Science Foundation Ireland (SFI) by funding {\'A}.~Vaca and F.~Milano under NexSys project, Grant No.~21/SPP/3756.}
\vspace{-0.6cm}
}%

\maketitle

\begin{abstract}
  This letter demonstrates how synthetic inertia can be obtained with the control of flexible discrete devices to keep the power balance of power systems, even if the system does not include any synchronous generator or conventional grid-forming converter.  The letter also discusses solutions to cycling issues, which can arise due to the interaction of uncoordinated discrete inertia controllers.  The effectiveness, dynamic performance, and challenges of the proposed approach are validated through simulations using modified versions of the WSCC 9-bus test system and of the all-island Irish transmission system.
\end{abstract}


\section{Introduction}

Synthetic inertia is defined as the controlled response of a generating unit to emulate the rotational energy exchange typical of synchronous machines \cite{56974154}.  Due to the high penetration in power systems of inverter-based resources, synthetic inertia has become essential for frequency stability, as it compensates for the decrease of physical inertia.  Foundational work, such as \cite{56974152} and \cite{56977468}, redefines system inertia and introduces virtual inertia strategies suited to modern grids.

Microgrids, often characterized by low inertia, benefit from synthetic inertia control models like those presented in \cite{56974162} and \cite{56974179}, which use inverter-based elements and energy storage to stabilize frequency.  At a larger scale, hybrid approaches such as distributed virtual inertia from supercapacitors \cite{56974170} and grid-scale energy storage systems \cite{56974181} offer innovative solutions for weak grids.  Other studies \cite{56977470} emphasize synthetic inertia's role in maintaining frequency stability in grids with high renewable penetration.  Moreover, inertia estimation methods reviewed in \cite{56974146} and \cite{56974147} are critical for adapting to fluctuating renewable energy sources while maintaining system stability.

Discrete Devices (DDs), also known as flexible resources, refer to controllable assets capable of operating in an on/off mode or at discrete power injection/consumption levels. These technologies include, but are not limited to, interruptible flexible loads, thermostatically controlled loads participating in demand response, prosumers, and inverter-based generators and storage systems that switch their power in discrete steps.

Over the past decade, DDs have emerged as a crucial component of power systems, providing services such as load shaving, and energy savings \cite{microflexibility}.  Their potential to contribute to frequency support, through stochastic control, of power systems has been widely investigated \cite{flexiLoads}.

While the literature demonstrates the importance of continuous controllers for frequency stability and resilience, there is limited research on how discrete devices can support a power system independently, i.e., without synchronous generators or conventional grid-forming converters.

Motivated by the theory of modeling and simulation, which states that continuous systems can be, in effect, considered a special case of discrete model events \cite{ziegler:00}, the novel contribution of this letter is a ``virtual swing control'' for discrete devices (DDs).  We show that the proposed control can guarantee the power balance of the overall system through the provision of a distributed discrete virtual inertia, as well as improve the dynamic performance of large power systems with the inclusion of conventional generation.  The letter also shows the challenges of discrete inertia control in standalone systems and offers potential solutions to these issues.

\section{Modeling}
\label{sec:method}

We consider the conventional model of power systems as hybrid automata, that is, as a set of differential-algebraic equations with the inclusion of discrete variables that define the behaviors of breakers and faults, as well as the connection status of discrete elements.  Formally, one has:
\begin{equation}
  \label{Eq:HDAE}
  \begin{aligned}
     \bfg x' &= \bfg{f}(\bfg{x}, \bfg{y}, \bfg{u}, \bfg{z}) , \\
     \bfg{0}_{\ny,1} &= \bfg{g}(\bfg{x}, \bfg{y}, \bfg{u}, \bfg{z}) ,
  \end{aligned}
\end{equation}
where $\bfg{f}$ are differential equations, $\bfg{g}$ are algebraic equations, $\bfg{x} = \bfg{x}(t) \in \mathbb{R}^{\nx}$ are state variables, e.g., machine rotor speeds, and $\bfg{y} = \bfg{y}(t) \in \mathbb{R}^{\ny}$ denote algebraic variables, e.g., bus voltage angles; $\bfg{u}$, $\bfg{u} \in \mathbb{R}^{n_u}$ are inputs, e.g., load variations; and $\bfg{z}$, $\bfg{z} \in \mathbb{R}^{\nz}$ are discrete variables, e.g., the status of breakers, tap ratios of under-load transformers and discrete devices (DDs).

In order to decide whether a DD has to change its power injection/consumption, we propose a ``virtual swing equation'', as follows:\color{black}
\begin{equation}
   \label{eq:dd}
  \begin{aligned}
  \delta' &= \omega - \omega_{\rm ref} , \\
  M \omega' &=  p_{\rm ref} - \tilde{p}_{e}(t, p_e) - D(\omega - \omega_{\rm ref}) ,  \\
  p'_{\rm ref} &=  \frac{1}{R} {\rm db}(\omega - \omega_{\rm ref}) - p_{\rm ref} , \\    
  p_e &= K_p v \sin(\delta - \theta) , 
\end{aligned}
\end{equation}
where $\delta, \omega, p_{\rm ref} \in \bfg x$ are the state variables of the virtual swing equation, which emulate the rotor angle and speed of a synchronous machine as well as the droop control of a turbine governor;  $p_{e} \in \bfg y$ is a virtual electrical power and $\tilde{p}_e \in \bfg z$ is the value of the actual power generation/consumption of the DD; $\theta$ is the phase angle of the terminal bus voltage as determined, for example, with a PLL; and $\omega_{\rm ref} \in \bfg u$ is the synchronous reference frequency; $M$ and $D$ are virtual inertia and virtual damping, respectively; $R$ is the droop coefficient; $v$ is the magnitude of the voltage at the device terminal bus; $K_p$ emulates the ratio between the internal emf and reactance, say $e/x$, of a synchronous machine; and $\rm db$ is a dead-band on the frequency error.

Equations \eqref{eq:dd} implement a ``modulation signal,'' specifically signal $p_e$, through a second-order model that behaves similarly to a synchronous machine.    Then the signal $p_e$ is utilized to decide the switching of the various elements of the flexible load.  The actual output power of the device,  $\tilde{p}_e$, is thus discrete, as follows:
\begin{equation}
  \label{Eq:Pdiscr}
  \tilde{p}_{e} = d_P(t, p_e) = h(t, p_e) \Delta p , \quad h = \{0, 1, \dots, n\} \, ,
\end{equation}
where $d_P(\cdot)$ is a discretization function of time and $p_e$ implementing the physical switching of the blocks $\Delta p$ of the device.  In practice, \eqref{Eq:Pdiscr} is automatically activated if $\omega$ is beyond the dead-band $\rm db$ in \eqref{eq:dd}, then $d_P(\cdot)$ determines the integer $h$ that approximates $p_e$ to the nearest multiple of discrete power packets $\Delta p$ included in each DD.  Note that $\tilde{p}_e$ is either that of generator or that of a flexible load, which can be considered a negative power injection.  Note also that each DD operates at regular but not synchronized time intervals and fully independently from all other devices.  This guarantees that the proposed virtual inertia control scales well and actually performs the better the higher of DDs participating to such a control.  Finally we assume that the reactive power injected/absorbed by each DD is either fixed, e.g., though a constant power factor, or regulated, e.g., in case of inverter-based resources.

\color{black}

In the following, we do not distinguish between DD generators and loads.  If the DD is a load, which is the most common case, then we assume that $\tilde{p}_e > 0$ is, in effect, implemented as a reduction of the load power consumption.  Finally, we assume that the power factor of DDs is constant, hence at a variation of $\tilde{p}_e$ corresponds a proportional variation of the reactive power.  Finally, each device evaluates \eqref{eq:dd} and \eqref{Eq:Pdiscr} at fixed intervals $\Delta t$ and, if necessary, changes its power injection $\tilde{p}_e$ only at the beginning of these intervals.

In this work we consider two scenarios: (i) standalone discrete devices (SDD), where discrete devices providing virtual inertia capability are the only devices in the system that are able to re-establish the power balance after a contingency; and (ii) complementary discrete devices (CDD), where discrete devices complement synchronous machines and or conventional gird-forming converters.  CDD scenario is the most likely scenario to happen in large transmission systems.  Yet, we show as a proof of concept that SDD is, in principle, at least, feasible and can be considered for small systems such as AC microgrids.

In the SDD scenario, properly calibrating the DDs sizes and operation times is essential.  These parameters must be carefully coordinated to ensure that the DDs have enough capacity to supply non-flexible power demand (adequacy) and be stable after a perturbation.  

The stability aspect requires further discussion.  Let the power balance equation of a system be:
\begin{equation}
  \label{Eq:Pbalance}
  \left | \sum{p_{{\rm gen}, i}} - \sum p_{{\rm DD}, k} - \sum{p_{{\rm load}, j}} - p_{\rm losses} \right | \le \varepsilon \, ,
\end{equation}
where $p_{{\rm gen}, i}$ is the active power injection of conventional generation (i.e., with continuous control); $p_{{\rm load}, j}$ is the active power consumption of every non-flexible load; $p_{{\rm DD}, k}$ is the combined power of DDs providing discrete inertia support; and $\varepsilon$ is a tolerance below which no switching of the discrete devices is activated.  If \eqref{Eq:Pbalance} is not satisfied, it is possible that the switching logic of the DDs can lead to cycling, that is, an unnecessary periodic switching on/off of DDs.  The cycling can be triggered if the size of available DDs is too big to satisfy \eqref{Eq:Pbalance} but also due to a combination of $\Delta t$ and $\Delta p$.  These aspects are illustrated in the case study.

\section{Case Study}
\label{sec:studycases}

This section evaluates the DDs described in Section \ref{sec:method} in two study cases: modified versions of the WSCC 9-bus test system and the Irish Power System.  All the simulations are performed using the power system analysis software tool Dome \cite{domePaper}.

\subsection{SDD Scenario: WSCC 9-Bus test system}

The WSCC 9-bus system, is a modified version based on \cite{AndersonFouad2003}, that includes 3 sets of PQ loads, connected to buses 5, 6, and 8, of 2, 0.9, and 1 pu(MW), respectively.  Loads at bus 5 are divided into three smaller groups of 1.1, 0.3, and 0.6 pu(MW).  Then, we assume that each bus of the grid represents the point of connection of a distribution system with the high-voltage transmission system.  Then we assume that the equivalent distribution system connected to each of these buses includes a high number of DDs.  Finally, for this case study, we assume that a subset of DDs has grid-forming capability to establish a reference frequency for the system and/or provide voltage support.  In total, $n = 300,000$ DDs are connected along the system buses equally split among all buses; the size of the DDs are uniformly distributed in the following set of values $\{10^{-2},10^{-3},10^{-4},10^{-5}\}$ pu(MW).  The operation of each DD is evaluated every $\Delta t = 1$ s.  The activation or deactivation of each DD is not synchronized, that is, DDs do not switch simultaneously.  Finally, to impose that only DDs support the power unbalance, the synchronous machines are modeled as constant PQ injections.

We test the dynamic performance of the system following 3 large perturbations:  (i) disconnection of the 1.1 pu(MW) PQ load of bus 5 at time $t = 15$ s; (ii) a 3-phase fault at $t = 20$ s, clearance time at $t = 22$ s, with a fault resistance of $10^{-3}$ pu($\Omega$); and (iii) a reconnection of the 1.1 pu(MW) PQ load of bus 5 at time $t = 35$ s.  In this scenario, the dead-band on the frequency error is set to zero.

\begin{figure}[htb]
    \centering
    \includegraphics[width=0.85\columnwidth]{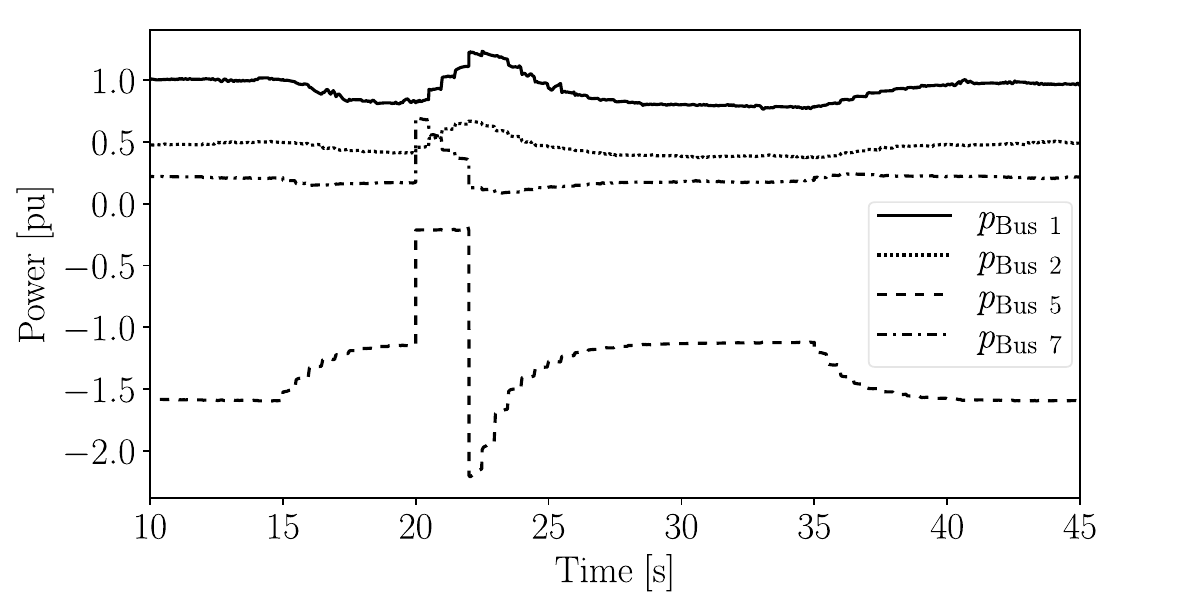}
    \caption{Power balance at buses 1, 2, 5 and 7 -- 9-bus WSCC System.}
    \label{fig:wscc2Power}
\end{figure}

Figure \ref{fig:wscc2Power} shows that, using $\Delta t = 1$ s, the system's power balance is fully sustained by DDs.

\begin{figure}[htb]
    \centering
    \includegraphics[width=0.85\columnwidth]{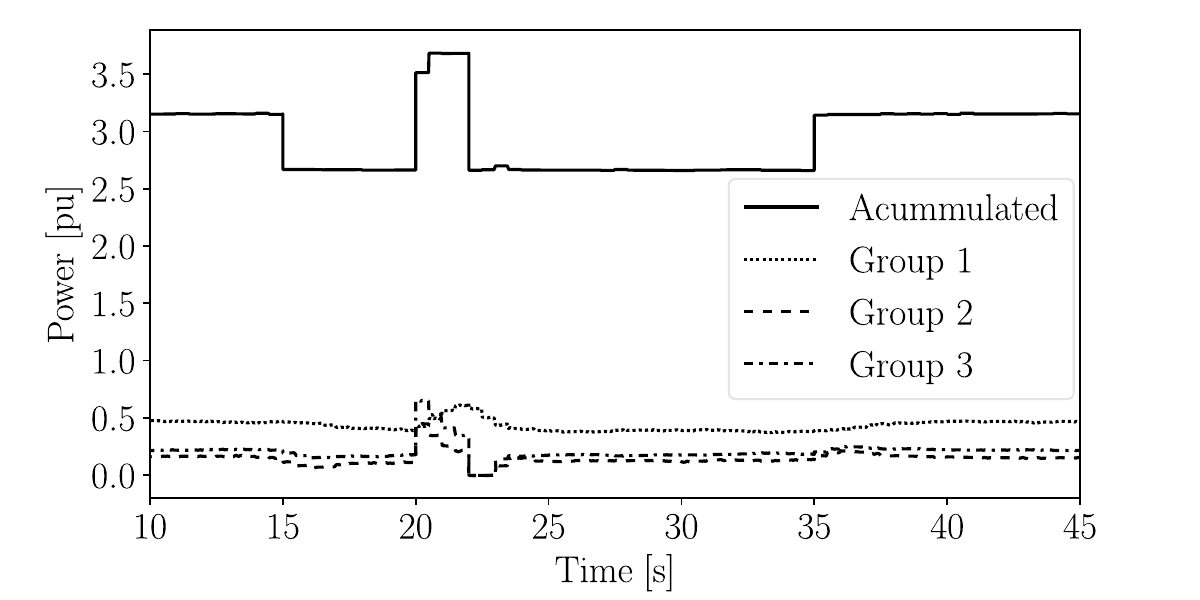}
    \caption{Active Power injection from groups of DDs and accumulated contribution of all DDs -- 9-bus WSCC System.}
    \label{fig:wscc2_1Power}
    \vspace{-2mm}
\end{figure}

Figure \ref{fig:wscc2_1Power} shows the active power injection from three randomly selected groups of discrete DDs and the aggregate contribution of all of them.  At $t = 15$ s, the system requires a reduction in power injection to regulate frequency; thus the groups respond with small numbers of individual DDs adjusting their output, resulting in a modest overall contribution.  Same small contribution is observed at $t = 35$ s, when the load is reconnected.  However, at $t = 20$ s, when the fault occurs, local frequency deviations trigger a rapid surge in DD power injections through the activation of a larger number of DDs.  The combination of high device granularity and decentralized control enables a near-continuous and fast response, effectively stabilizing the system.

\begin{figure}[htb]
    \centering
    \includegraphics[width=0.85\columnwidth]{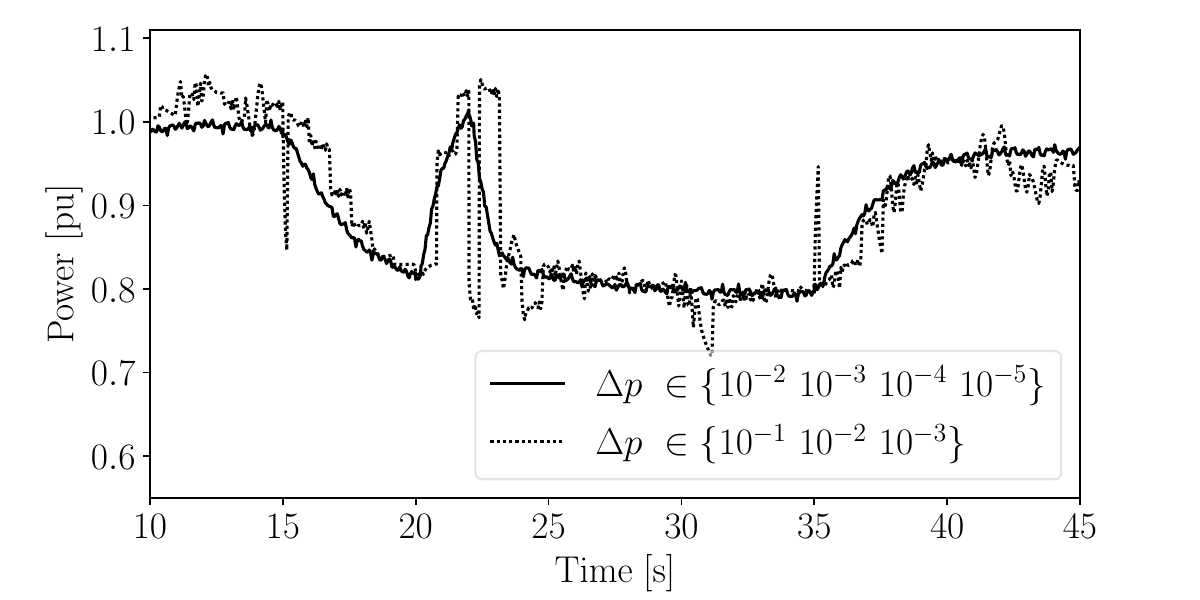}
    \caption{Power injection at bus 3 for different DDs' size combination -- 9-bus WSCC System.}
    \label{fig:wsccjumps}
    \vspace{-2mm}
\end{figure}

When this combination is not adequately achieved, DDs react with one another, causing cycling behavior since \eqref{Eq:Pbalance} is not satisfied. This phenomenon is illustrated in Fig. \ref{fig:wsccjumps}, which shows the effect of total power injection on a bus for two combinations of DD sizes. One trivial solution to this issue is to use only small-sized DDs.  However, this would require a much larger number of devices to meet the system's power demands.  Instead, a proper balance among device sizes, granularity, and operational characteristics of the DDs ensures adequate functioning under SDD.

\begin{figure}[htb]
    \centering
    \includegraphics[width=0.85\columnwidth]{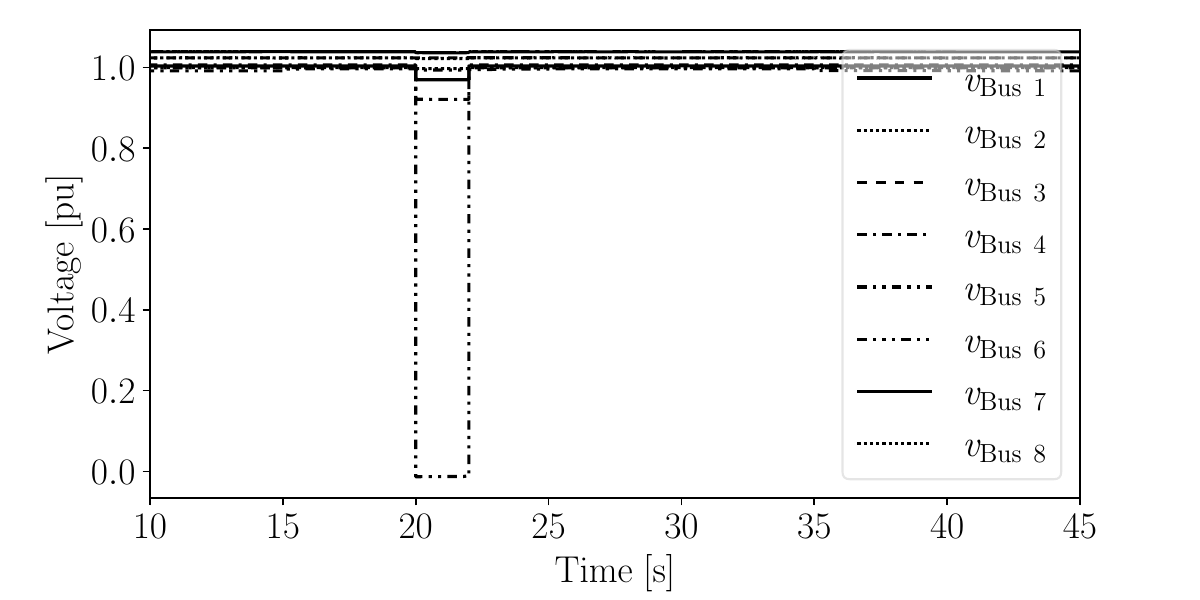}
    \caption{Voltages on the busses -- 9-bus WSCC System.}
    \label{fig:wscc3Voltage}
    \vspace{-2mm}
\end{figure}

Figure \ref{fig:wscc3Voltage} shows that system voltages remain within acceptable operational levels.  Following the clearance of the 3-phase fault occurred at bus 5, DDs effectively contributes to support the voltages at other buses by rapidly increasing active and reactive power injection, as in Fig.~\ref{fig:wscc2Power}.  In turn, the switching logic works because the number of DDs is sufficiently high that the overall effect of the DDs in this case study approximates the behavior of a ``distributed infinite bus.''

\subsection{CDD Scenario: All-island Irish transmission system}

We use a dynamic model of the all-island Irish transmission system to demonstrate the performance of DDs in a real-world grid.  The system's model is based on the public information of the transmission system operators of the island, EirGrid and SONI, and on \cite{frequency-dependent_2019}.  The model includes 1479 buses, 1851 lines, 5 conventional power plants, and 302 wind power plants with a caseload of 1.8 GW.  In all cases, DDs evaluate their participation every $\Delta t = 1$ s, $n = 100,000$ DDs, and the evaluation time of each DD is randomized in this interval.

\begin{figure}[htb]
    \centering
    \includegraphics[width=0.85\columnwidth]{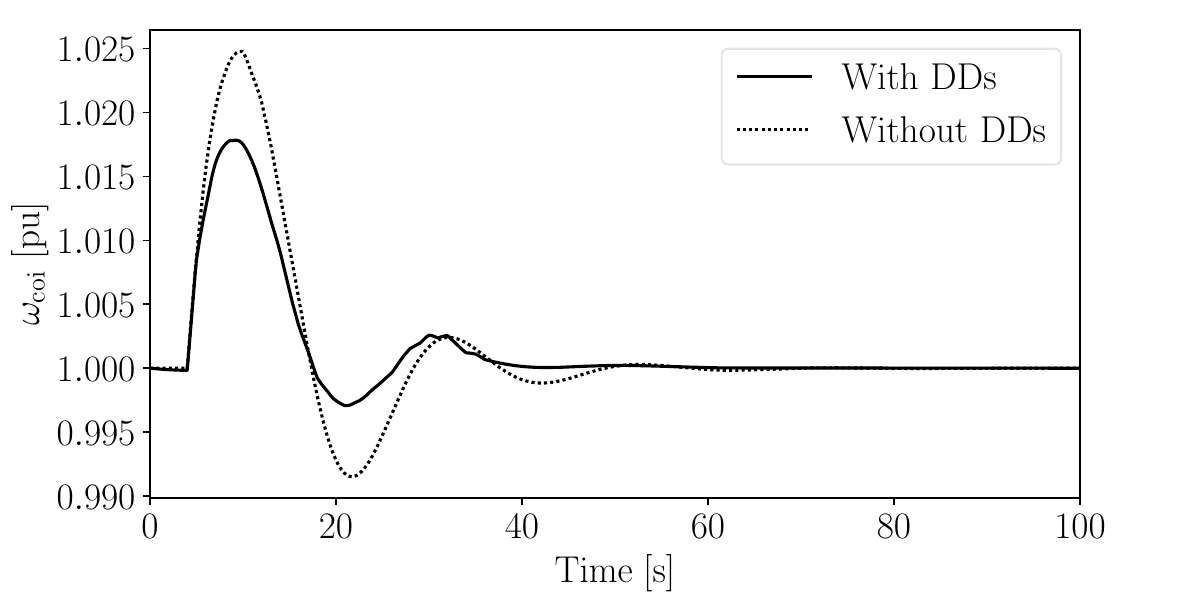}
    \caption{Frequency support from DDs -- Irish System.}
    \label{fig:supportIrish}
    \vspace{-2mm}
\end{figure}

The contingency is a disconnection of an equivalent load of 7\% of the system.  Figure \ref{fig:supportIrish} shows the dynamic performance of the model of the Irish system as a base case and how its dynamic performancs is improved through CDD support. 

\begin{figure}[htb]
    \centering
    \includegraphics[width=0.85\columnwidth]{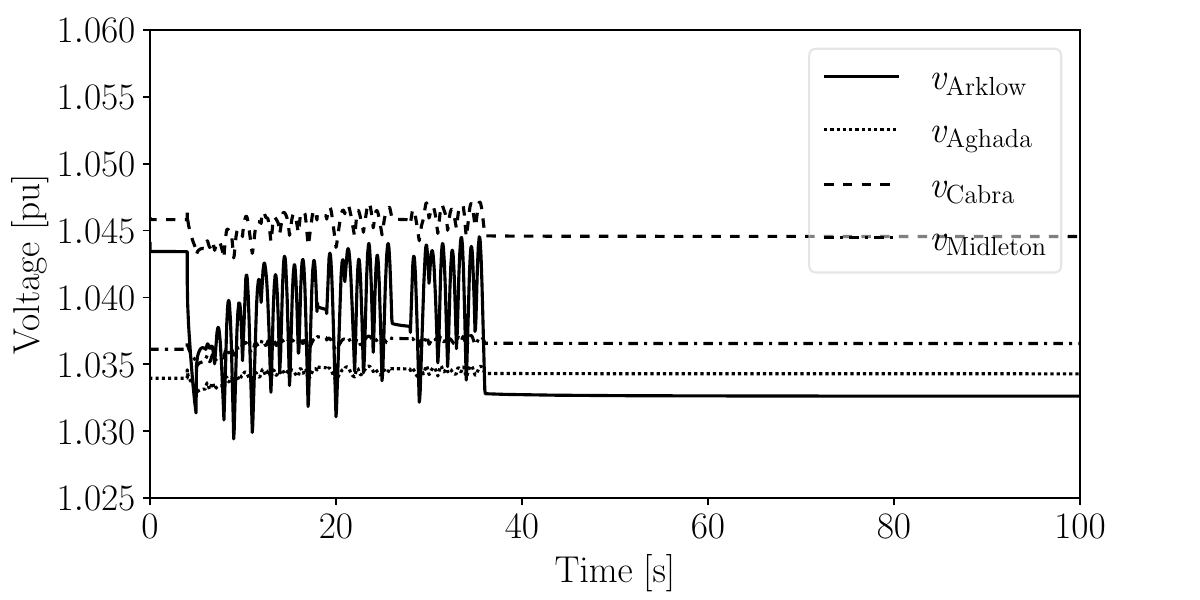}
    \caption{Voltage on representative busses -- Irish system.}
    \label{fig:voltagesupportIrish}
    \vspace{-2mm}
\end{figure}

\begin{figure}[htb]
    \centering
    \includegraphics[width=0.85\columnwidth]{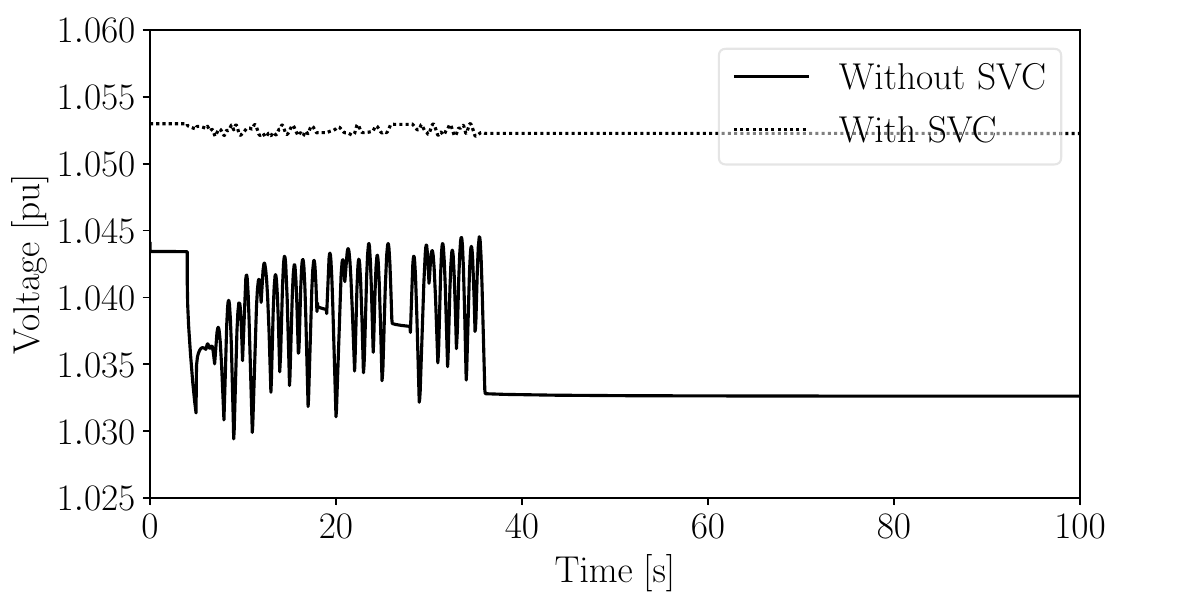}
    \caption{Correction of voltage oscillation -- Bus Arklow.}
    \label{fig:voltagesupportIrishSVC}
    \vspace{-2mm}
\end{figure}

Figure \ref{fig:voltagesupportIrish} shows that the switching of the DDs triggers slight voltage oscillations. The oscillations are relatively small due to the high granularity and balanced sizes of the DDs. They can be mitigated by connecting a dedicated device, as shown in Fig. \ref{fig:voltagesupportIrishSVC}, where an SVC is connected to bus Bedford.

\section{Conclusions}
\label{sec:conclusion}

This study demonstrates the potential of flexible discrete devices to provide synthetic inertia to power systems.  DDs are shown to effectively manage frequency and voltage stability autonomously by injecting discrete power packets.  The paper also shows that proper calibration of the DDs control can eliminate cycling effects, typical of the interaction of discrete controllers.  The application of this approach to the WSCC 9-bus and the all-island Irish transmission systems confirms that DDs can offer scalable, adaptive support to meet dynamic stability requirements in power grids.

Future research will focus on integrating DDs with other fast-response technologies, potentially broadening DDs' application in various power system contexts.


\vfill

\begin{thebibliography}{10}
\providecommand{\url}[1]{#1}
\csname url@samestyle\endcsname
\providecommand{\newblock}{\relax}
\providecommand{\bibinfo}[2]{#2}
\providecommand{\BIBentrySTDinterwordspacing}{\spaceskip=0pt\relax}
\providecommand{\BIBentryALTinterwordstretchfactor}{4}
\providecommand{\BIBentryALTinterwordspacing}{\spaceskip=\fontdimen2\font plus
\BIBentryALTinterwordstretchfactor\fontdimen3\font minus
  \fontdimen4\font\relax}
\providecommand{\BIBforeignlanguage}[2]{{%
\expandafter\ifx\csname l@#1\endcsname\relax
\typeout{** WARNING: IEEEtran.bst: No hyphenation pattern has been}%
\typeout{** loaded for the language `#1'. Using the pattern for}%
\typeout{** the default language instead.}%
\else
\language=\csname l@#1\endcsname
\fi
#2}}
\providecommand{\BIBdecl}{\relax}
\BIBdecl

\bibitem{56974154}
R.~Eriksson, N.~Modig, and K.~Elkington, ``Synthetic inertia versus fast
  frequency response: a definition,'' \emph{IET Renewable Power Generation},
  vol.~12, no.~5, pp. 507--514, 2018.

\bibitem{56974152}
M.~Rezkalla \emph{et~al.}, ``Comparison between synthetic inertia and fast
  frequency containment control based on single phase {EVs} in a microgrid,''
  \emph{Applied Energy}, vol. 210, pp. 764--775, 2018.

\bibitem{56977468}
P.~Tielens and D.~Van-Hertem, ``\BIBforeignlanguage{en}{The relevance of
  inertia in power systems},'' \emph{\BIBforeignlanguage{en}{Renewable and
  Sustainable Energy}}, vol.~55, pp. 9--1, 2016.

\bibitem{56974162}
S.~Nema, V.~Prakash, and H.~Pandžić, ``Adaptive synthetic inertia control
  framework for distributed energy resources in low-inertia microgrid,''
  \emph{IEEE Access}, vol.~10, pp. 54\,969--54\,979, 2022.

\bibitem{56974179}
U.~Bose, S.~Chattopadhyay, C.~Chakraborty, and B.~Pal, ``A novel method of
  frequency regulation in microgrid,'' \emph{IEEE Transactions on Industry
  Applications}, vol.~55, no.~1, pp. 111--121, 2019.

\bibitem{56974170}
M.~Saeedian, B.~Pournazarian, S.~Taheri, and E.~Pouresmaeil, ``Provision of
  synthetic inertia support for converter-dominated weak grids,'' \emph{IEEE
  Systems Journal}, vol.~16, no.~2, pp. 2068--2077, 2022.

\bibitem{56974181}
J.~Mitra and N.~Nguyen, ``Grid-scale virtual energy storage to advance
  renewable energy penetration,'' \emph{IEEE Transactions on Industry
  Applications}, vol.~58, no.~6, pp. 7952--7965, 2022.

\bibitem{56977470}
J.~Liang, H.~Fan, Z.~Deng, J.~Zhu, L.~Yu, T.~Li, P.~Luo, Z.~Gao, and Y.~Wang,
  ``Generic synthetic inertia scheme for voltage source inverters,'' \emph{IET
  Renewable Power Generation}, vol.~17, no.~3, pp. 696--705, 2 2023.

\bibitem{56974146}
K.~Prabhakar, and P.~Padhy, ``Inertia estimation in modern power
  system: A comprehensive review,'' \emph{Electric Power Systems Research}, 2022.

\bibitem{56974147}
K.~Ratnam, K.~Palanisamy, and G.~Yang, ``Future low-inertia power systems:
  Requirements, issues, and solutions - a review,'' \emph{Renewable and
  Sustainable Energy Reviews}, vol. 124, p. 109773, 2020.

\bibitem{microflexibility}
S.~Chatzivasileiadis \emph{et al.}, ``Micro-flexibility: Challenges for power system modeling and control,'' \emph{Electric Power Systems Research}, vol.~216, p.~109002, 2023.

\bibitem{flexiLoads}
J.~McMahon, T.~Kërçi, and F.~Milano, ``Combining Flexible Loads with Energy Storage Systems to provide Frequency Control,'' in \emph{IEEE PES Innovative Smart Grid Technologies - Asia (ISGT Asia)}, 2021

\bibitem{ziegler:00}
B.~Ziegler, H.~Praehofer, and T.~Kim, \emph{Theory of Modeling and
  Simulation}.\hskip 1em plus 0.5em minus 0.4em\relax London, UK: Academic
  Press, 2000.

\bibitem{domePaper}
F.~Milano, ``A {Python}-based software tool for power system analysis,'' in
  \emph{IEEE PES General Meeting}, 2013, pp. 1--5.

\bibitem{AndersonFouad2003}
P.~Anderson and A.~Fouad, \emph{Power System Control and Stability},
  2nd~ed.\hskip 1em plus 0.5em minus 0.4em\relax IEEE Press/Wiley-Interscience,
  2003.

\bibitem{frequency-dependent_2019}
\BIBentryALTinterwordspacing
F.~Milano and A.~Ortega, ``Frequency-{Dependent} {Model} for {Transient}
  {Stability} {Analysis},'' \emph{IEEE Transactions on Power Systems}, vol.~34,
  no.~1, pp. 806--809, 2019.
\BIBentrySTDinterwordspacing

\end{thebibliography}
\end{document}